# Modelling Dynamical Fluorescent Micro Thermal Imaging of the Heat Diffusion in the $La_5Ca_9Cu_{24}O_{41}$ Spin Ladder Compound.

E.I. Khadikova*[1], M.Montagnese[2], F. de Haan[1], P.H.M. van Loosdrecht[1,2].
[1]Zernike Institute for Advanced Materials, University of Groningen, the Netherlands.
[2]II. Physics Institute, University of Cologne, Köln, Germany.
*Corresponding author: Nijenborgh 4, 9747 AG Groningen, the Netherlands, e.i.khadikova@rug.nl

**Abstract:** The dynamical fluorescent micro-thermal imaging (FMI) experiment has been used to investigate the phonon-magnon interaction in the 1D Heisenberg antiferromagnet $La_5Ca_9Cu_{24}O_{41}$. This material shows highly anisotropic heat conductivity due to the efficient magnetic heat transport along the spin ladders in the compound carried by magnetic excitations (magnons). To extract information on the phonon-magnon interaction we modelled the dynamic heat transport experiment using a two temperature model approach and taking both the crystal as well as the europium (III) thenoyltrifluoroacetonate/deuterated Poly(methyl methacrylate) fluorescent heat imaging layer into account. The simulations are carried out by the finite element method using COMSOL Multiphysics Heat Transfer. The results of the numerical calculations are crucial to the data analysis of the experimental studies.

**1. Introduction**

The thermal management of electronic devices and systems remains an important issue in view of the continuing miniaturization in the semiconductor technology. Low dimensional quantum magnets are good candidate materials for heat extraction structures, since they have both high thermal conductivity as well as electrical insulating behavior near room temperature. [1] In addition these materials show a high heat transport anisotropy which potentially allows for novel, more efficient device designs. [2] The fact that in a heat transport experiment one has only indirect access to the magnetic excitation (through phonon-magnon coupling [3,4]) allows one to study the spin-lattice coupling in these materials through dynamic heat transport experiments. Understanding spin-lattice coupling is not only important in these materials, it plays also an important role in many other phenomena observed in condensed matter physics, such as the Spin-Seebeck effect and the Spin-Peierls instability. Phonon-magnon coupling is one of the aspects which determine the scattering of magnons and their lifetime. Theoretical calculations of the spin wave lifetime have been performed over 40 years, a review can be found in [6]. However experimental techniques available to date are indirect and give contradictory results [7, 8], so there is a need for a novel approach as discussed here.

The crystalline structure of the one dimensional Heisenberg antiferromagnet $La_5Ca_9Cu_{24}O_{41}$ constitutes the spin ladders with strong antiferromagnetic exchange interaction between the spins of the copper ions along the lags and rungs of the ladders. The magnetic excitations show a strong dispersion along the ladder direction and a fairly flat one in the two other directions. This leads to both the high diffusivity of the magnetic excitations as well as to the strong anisotropy of the magnetic heat diffusion. [1]

This work deals with the modelling of the dynamical fluorescent micro-thermal imaging experiment on materials where different excitations contribute to the thermal heat transport. This type of heat diffusion will be modeled using a two temperature (2T) model as in [4], with one temperature corresponding to the phonons, and one to the magnons in the system. In addition the influence of the fluorescent europium thenoyltrifluoroa-cetonate (EuTTA) / deuterated Poly-methyl methacrylate(PMMA) layer probing layer on the experimental temperature profile of the heat diffusion in the investigated compound will be addressed.

**2. The dynamical fluorescent micro-thermal imaging experiment.**

The results of the numerical simulations are to be compared with the experimental results obtained by the dynamical FMI technique. For the experiments a platelet of the the $La_5Ca_9Cu_{24}O_{41}$ compound is cut from a large crystal to have the spin ladders parallel to the platelet plane (ac crystallographic plane). Subsequently a layer of PMMA/EuTTA is spin-coated on the platelet. This sample is mounted on a microscope where a short (20μs) focused 488nm laser pulse is used to create a hot spot on the sample. The temporal temperature profile evolution is subsequently measured by exciting the PMMA/EuTTA layer using a short defocused 370 nm laser pulse (20μs) at various times after the creation of the hot spot. The microthermal images are detected using a 20 x magnification objective and a CCD camera, and converted into temperature profiles using a calibration curve. For the experiments we used two different PMMA/EuTTA thicknesses – 0.4μm and 2μm, in order to investigate the contribution of the polymer layer to the imaging of the heat diffusion. The resulting spatial temperature profiles were fitted by two dimensional (2D) Gaussians corresponding to the heat diffusion in the sample and the polymer layer; the temporal evolution of the experimental widths of the Gaussians are shown in Fig.2.

## 3. Modelling of the multicomponent heat transport
### 3.1 Heat Transport Model

The dynamical heat transport was modelled using two different approaches. First one, called here the one temperature (1T) model is based on the numerical calculations of the standard time dependent heat equation, with the heating laser pulse applied to the sample directly:

$$c_{tot}\partial_t T = \nabla(k_{tot}\nabla T) + Q \quad (1)$$

where $c_{tot}$, $T$, $k_{tot}$ are the specific heat, temperature and thermal conductivity for the lattice and magnon subsystems, and Q is a laser heat source.

The second approach is based on the two temperature (2T) model, characterized by two different contributions to the heat transport originating from the magnon and phonon subsystems resolved by the fully-coupled solver.

The corresponding equations to be solved using the fully coupled solver are:

$$\begin{cases} c_l\partial_t T_l = \nabla(k_l\nabla T_l) - g(T_l - T_m) + Q \\ c_m\partial_t T_m = \nabla(k_m\nabla T_m) - g(T_m - T_l) \end{cases} \quad (2)$$

where $c_l$, $k_l$, $T_l$ and $c_m$, $k_m$, $T_m$ are the specific heat, thermal conductivity and temperature of the phonon and magnon systems, respectively, and g is the coupling constant between the magnons and the phonons:

$$g = \frac{c_m c_l}{c_m + c_l}\frac{1}{\tau_{ml}} \quad (3)$$

where $\tau_{ml}$ is the phonon-magnon thermalization time. The 3D model of the dynamical fluorescent micro-thermal imaging (FMI) experiment is represented by the bulk heat propagation in time of the heat pulse through the sample covered by a polymer layer with two different thicknesses – 0.4 μm and 2 μm.

Dimensions of the sample are determined by the spatial extend of the heat diffusion within the studied time interval, and are 20 μm x 2000 μm x 60 μm for the one temperature(1T) model, and 20 μm x 1000 μm x 60 μm for the 2T model.

For the volume (9μm x 300μm x 40μm) around the initial hot spot we used a finer spatial and temporal mesh to model the more rapid changes in this area.

The heating laser pulse is represented as a Gaussian function in space with σ = 2.3 μm and a block function in time with duration of 20μs. The energy of the laser pulse is taken as E=0.02μJ per pulse, corresponding to the experimental conditions. The initial temperature of all domains was 293.15K. The values of the thermal conductivity, specific heat and density of the spin ladder compound were taken from [3] and are shown in the Table 1.

The absorption of the laser pulse by the polymer layer (<4%) has been ignored in the simulations and the heat source has been applied directly to the spin-ladder compound. The external surfaces were modeled as insulated from the environment; the temperature gradients across these surfaces were set to zero. The internal boundaries conditions were assumed to be thermally continuous, implying an equality of the temperatures and heat fluxes at both sides of the boundary of the two media.

The mesh used for the simulation by both models was free triangular for coarse and finer mesh domains with a maximum element growth rate 1.5, curvature factor 0.6 and resolution of the narrow regions 0.5, as shown in Fig.1. The size of the domains for 1Temperature model and 2Temperature model respectively was varied from 50µm and 200µm to 10µm and 10µm for the coarser mesh and from 1.5µm and 30µm to 0.5µm and 1µm for the finer mesh. Complete mesh consisted of a maximum of 1869967 domain elements with 2504325 degrees of freedom to be solved for.

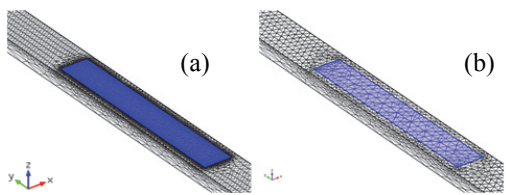

**Figure 1**. The space mesh of the geometry of the 1T (a) and the 2T (b) models, blue areas corresponds to the finer mesh around the hot spot.

The time mesh was explicitly specified by an exponential distribution during the application of the laser pulse. A linear distribution with step 1.25µs have been used for the rest of the heat dynamics. The temperature distribution in the model sample was calculated for 249 time delays in total covering a time span of 0 to 300µs for each thickness of the polymer layer.

### 3.2 Simulation Results

The temperature distributions of a xy plane cross section of the polymer layer for the different time delays averaged along the z axis are shown in Fig.2. Each profile has been fitted by sum of two Gaussian functions (as has been done for the experimental data):

$$F(x) = H_1 e^{-0.5\left(\frac{x-x_{01}}{\sigma_1}\right)^2} + H_2 e^{-0.5\left(\frac{x-x_{02}}{\sigma_2}\right)^2} \quad (4)$$

Where widths $\sigma_1$ and $\sigma_2$ corresponds to the effective thermal diffusivities of the sample and the polymer layer, respectively; and amplitudes, $H_1$ and $H_2$, gives the relative importance of the two transport channels.

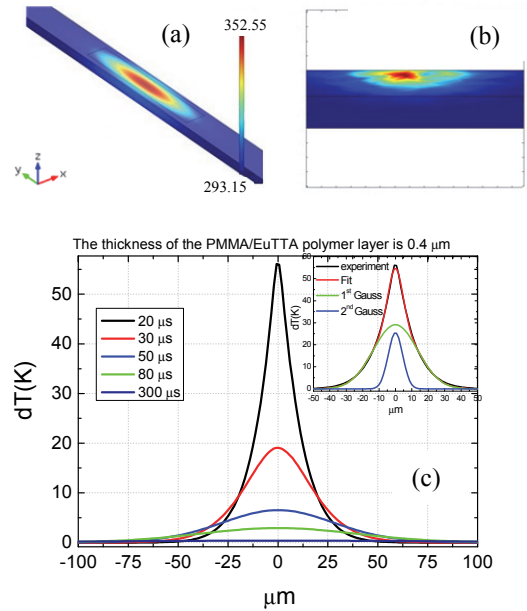

**Figure 2**. Example of the data analysis. (a) 3D temperature distribution of the sample corresponding to the anisotropic heat transport along the spin ladders for 1T model approach.(b) 2D temperature distribution of the XY plane cross section of the sample. (c) 1D temperature distribution corresponding to the average of 2D temperature distribution of the polymer layer along the z axis for five different time delays. Insert: Fitting of the 1D temperature distribution by two Gaussian functions.

A graphical representation of the numerical results using the 1T model is shown in Fig.3. Plots (a) and (b) show the widths of the fitted Gaussians, for two different thicknesses of the polymer layer extracted from the temperature profile of the cross-sections parallel to the spin ladders and perpendicular to the heated surface, as a function of delay time.

The simulation and experimental results are collated also to the diffusion functions – $\sqrt[2]{2Dt + \sigma_0^2}$, where D is diffusivity of the $La_5Ca_9Cu_{24}O_{41}$ and $PMMA/EuTTA$ and $\sigma_0$ is the width of the laser pulse, as used in the model.

A graphical representation of the numerical calculations using the 2T model is shown in the Fig.4.

Plots (a), (b), (c), and (d) show an example of the 3D distribution of the phonon (a,b) and magnon (c,d) temperature directly after the laser pulse Δt = 20µs (a, b) and for the last calculated time delay Δt = 300µs (c, d).

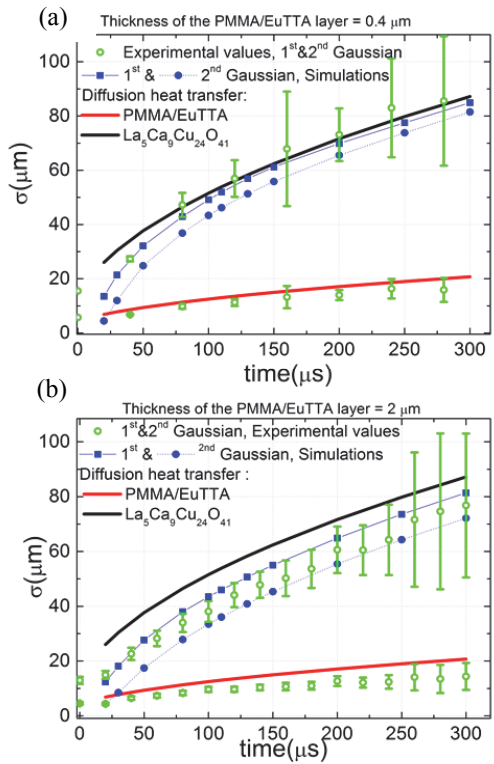

**Figure 3.** 2T model simulations results. Blue lines: the time dependence of widths of the 1st and 2nd Gaussians extracted from the 2D temperature profile of the polymer layer with two different thicknesses – 0.4μm (a) and 2μm (b), along the spin ladders. Red and black lines: the time dependence of the Gaussian width corresponding to the heat diffusion in PMMA/EuTTA layer and spin ladder compound, respectively, directly calculated from the thermal diffusivities (see the appendix). Green lines: time dependence of the 1st and 2nd Gaussians extracted from the experimental data with the corresponding confidence intervals.

For both the 1T and 2T models we observed a smaller width of the 1st Gaussian for larger thickness of the polymer layer, indicating smaller effective thermal conductivities of the sample for larger thickness of the PMMA/EuTTA layer. The experimental data is in good agreement with the width of the 1st Gaussian width for 1T model simulations, whereas, the width of the 2nd Gaussian is much larger compared to the experimental results. This is not unexpected since the 1T model cannot capture the multicomponent transport happening in the spin-ladder compounds. The resulting widths of the 1st Gaussians for 2T model simulations appear to be much smaller than experimental data, whereas 2nd Gaussian widths are comparable with the experiment values. The 2T approach was used with slightly coarser mesh and slightly larger time steps in order to reduce the amount of used computer resources; which could lead to insufficient coupling of the two physics corresponding to phonon and magnon systems in the computing process.

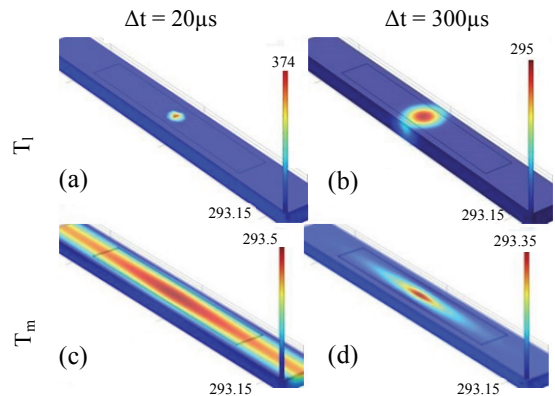

**Figure 4.** 2T model simulations results. (a, b) Phonon 3D temperature distribution directly after the applying of the heat laser pulse and at the last calculated time delay ($\Delta t = 300\mu s$), respectively, for the 2T model approach. (c,d) Magnon 3D temperature distribution for the same time delays for the 2T model approach.

### 4. Conclusions

Present study shows a strong correlation between the thickness of the polymer layer and observed thermal conductivity of the sample for both used approaches. This result can be used for the data analysis of the dynamical-FMI experimental studies in order to exclude the heat transport in polymer layer from the experimental results by fitting the temperature profiles with a sum of two Gaussian functions. The 2T approach shows not efficient phonon-magnon coupling, as can be seen from a comparison with experimental data. The application of the 2Temperature model requires further modelling.

## 5. References

1. C. Hess et al., Magnon heat transport in (Sr, Ca, La)$_{14}$Cu$_{24}$O$_{41}$, Physical Review B, vol. 64, 184305 (2001).
2. M. Otter et al., Optical probing of anisotropic heat transport in the quantum spin ladder Ca$_9$La$_5$Cu$_{24}$O$_{41}$, Int. J. Heat and Mass Transfer 55, 2531 (2012)
3. S.I. Anisimov et al., Electron Emission from Metal Surface Exposed to Ultrashort Laser Pulses, *Sov. Phys. JETP*, Vol. 39, 375-377 (1974).
4. M. Montagnese et al., Phonon-magnon interaction in low dimensional quantum magnets observed by dynamic heat transport measurements, Phys. Rev. Letter, 110, 147206 (2013).
5. A.V. Sologubenko et al., Phys. Rev. Lett. 84, 2714 (2000).
6. I.F.I. Mikhail et al., Exact and model operators for magnon–phonon interactions in antiferromagnets, Physica B 406, 508–515 (2011).
7. Z. Yamani et al., Neutron scattering study of the classical antiferromagnet MnF2: a perfect hands-on neutron scattering teaching course, Can. J. Phys. 88: 771–797 (2010)
8. D. J. Sanders et al., Effect of magnon-phonon thermal relaxation on heat transport by magnons, Phys. Rev.B 15, 1489 (1977).

## 6. Acknowledgements

This work was supported by the European Commission through the ITN Marie Curie LOTHERM, Contract Number: PITN-GA-2009-238475.

## 7. Appendix

**Table 1:** 2T Model parameters using in the numerical simulations.

| Quantity | $La_5Ca_9Cu_{24}O_{41}$ | PMMA / EuTTA layer | Dimensions |
|---|---|---|---|
| $k_l$ | 1* | 1.19* | WK$^{-1}$m$^{-1}$ |
| $k_m$ | 34.5* | - | WK$^{-1}$m$^{-1}$ |
| $c_l$ | 27.42[4] | 1466 | JK$^{-1}$kg |
| $c_m$ | 522.913[4] | - | JK$^{-1}$kg |
| $\rho$ | 5469.36 | 1190 | kg m$^{-3}$ |
| $\tau_{ml}$ | 4e-4[4] | - | sec |

* dynamical FMI measured values